\documentclass[aps,prd,groupedaddress]{revtex4-1}

\usepackage{graphicx}
\usepackage{dcolumn}
\usepackage{bm}
\usepackage{amsmath}
\usepackage{amssymb}
\usepackage{color}
\usepackage{ulem}
 
\def\mbi#1{\mbox {\bfseries\itshape #1}}

\newcommand{\aap}{\rm Astron.~Astrophys.~}

\newcommand{\apjl}{\rm Astrophys.~J.~Lett.~}
\newcommand{\apjs}{\rm Astrophys.~J.~Supp.~}

\newcommand{\jcap}{\rm JCAP~}

\newcommand{\physrep}{\rm Phys.~Rep.~}

\begin{document}


\title{Evolution of the cosmic matter density field with a primordial magnetic field}

\author{Dai G. Yamazaki$^{1,2}$}
 \email{yamazaki.dai@nao.ac.jp}
\affiliation{%
$^{1}$National Astronomical Observatory of Japan, Mitaka, Tokyo 181-8588, Japan}%
\affiliation{%
$^{2}$Ibaraki University, 2-1-1, Bunkyo, Mito, 310-8512, Japan}%
\date{\today}

\begin{abstract}
A cosmological magnetic field affects the time evolution of the cosmic matter density field. The squared Alfven velocity of the cosmic fluid is proportional to an ensemble average energy density of a primordial magnetic field (PMF), and it prevents the matter density field from collapsing in the horizon scale.
The matter-radiation equality time also is delayed by the presence of an ensemble average energy density of a PMF. The ensemble average energy density of the PMF also affects the matter power spectrum (MPS) through the Meszaros effect and the potential decay. 
Since the ensemble average energy density of the PMF is not a first order perturbation but a zero order source in the linear perturbation equations for the cosmology, to correctly understand the overall effects of the PMF on the MPS, we should significantly revise previous approaches to research for the MPS with the PMF by considering the both effects of the zero and first order sources from the PMF in the linear perturbation theory.
We apply the effects of the zero order sources from the PMF to theoretical computations of the MPS for the first time. We also analyze the overall PMF effects on the MPS. The CMB polarizations are affected the weak lensing. The weak lensing is determined by the MPS. Therefore, we have to consider the zero order sources of the PMF to gain a correct understanding not only of the MPS but also the CMB polarization.
\end{abstract}
\maketitle 
\section{\label{sec:introduction}Introduction}
Observations of the Faraday rotation and synchrotron emission \cite{2004IJMPD..13.1549G,Wolfe:1992ab,Clarke:2000bz,Xu:2005rb,2010Sci...328...73N,2012ApJ...747L..14V,2012A&ARv..20...54F} show that the present of the magnetic field of 1 $\mu$G ($=10^{-6}$ G) order turns out to be in the length between the typical sub-galaxy to cluster scale. 
One of the most famous ideas to explain these magnetic fields is that homogeneous and stochastic magnetic fields of 1 nG ($=10^{-9}$ G) order are generated before the recombination, and these evolve the observed magnetic fields by the isotropic collapses of density fields in the early Universe.  \cite{Grasso:2000wj,2010AdAst2010E..80Y,2011PhR...505....1K,2012PhR...517..141Y}.

A primordial magnetic field (PMF), which is expected to exist in the early Universe as mentioned above, affects the cosmic microwave background (CMB) and the matter power spectrum (MPS). 

Many groups analyze the PMF effects on the CMB
\cite{1996ApJ...469....1K,1998PhRvD..58l3004D,Subramanian:1998fn,Jedamzik:1999bm,Durrer:1999bk,Grasso:2000wj,Mack:2001gc,2002PhRvD..65h3502P,Subramanian:2002nh,2004PhRvD..70f3003S,Giovannini:2004aw,Kosowsky:2004zh,2004IJMPD..13..391G,2004PhRvD..69f3006C,Kahniashvili:2005xe,2005NewAR..49...79K,Yamazaki:2005yd,Kahniashvili:2005xe,2005PhRvD..72j3003G,Dolgov:2005ti,2005PhRvD..72j3003G,Kahniashvili:2006hy,Giovannini:2006kc,Giovannini:2007qn,2008PhRvD..77d3005Y,Paoletti:2008ck,2008nuco.confE.239Y,Sethi:2008eq,Kojima:2008rf,2008PhRvD..78f3012K,Giovannini:2008aa,Finelli:2008xh,2008PhRvD..77l3001G,2008PhRvD..78f3012K,Paoletti:2008ck,2010AdAst2010E..80Y,2010JCAP...05..022B,2011PhR...505....1K,2011PhRvD..84l3006Y,2011PhRvD..83b3006K,2012PhR...517..141Y,2013PhRvD..88h3515B,2014PhRvD..89j3528Y} and constrain the PMF by the CMB observations \cite{Barrow:1997mj,Jedamzik:1999bm,Yamazaki:2004vq,Lewis:2004ef,Yamazaki:2006bq,Yamazaki:2006ah,Yamazaki:2008bb,Kahniashvili:2008hx,2009JCAP...07..041K,2010PhRvD..81b3008Y,2010PhRvD..82h3005K,2011PhRvD..83l3533P,2011PhRvD..84d3530P,2012PhRvD..86d3510S,2012PhRvL.108w1301T,2013PhRvD..88j3011Y,2013PhLB..726...45P,2014PhRvD..90h3004K,2014PhRvD..90j3002S,2014JCAP...01..009K}.
Some groups also analyze the PMF effects on the bispectrum or/and non-Gaussianity of the CMB \cite{Brown:2005kr,Caprini:2009vk,Seshadri:2009sy,2010JCAP...08..025C,2010PhRvD..82l1302S,2011PhRvD..83l3003S,2012JCAP...03..041S}.
They mention that the PMF increases the CMB on the smaller scales. They also constrain the upper bound of the PMF strength as a 1 nG order by the CMB observations.

The PMF effects on the cosmological density fields or the matter power spectrum (MPS), also, are studied by the many authors \cite{1996ApJ...468...28K,2003JApA...24...51G,Sethi:2003vp,Sethi:2004pe,2006PhRvD..74l3518Y,Yamazaki:2008bb,2008nuco.confE.239Y,2010PhRvD..81j3519Y,2010PhRvD..82h3005K,2012SSRv..166....1R,2012JCAP...11..055F,2013ApJ...770...47K,2013ApJ...762...15P}. 
References \cite{1996ApJ...468...28K, 2003JApA...24...51G,Sethi:2003vp,Sethi:2004pe} groundbreaking analytical research on the effects of the first order perturbations source from PMFs on the MPS. Then Refs. \cite{2006PhRvD..74l3518Y,Yamazaki:2008bb,2008nuco.confE.239Y,2010PhRvD..81j3519Y} develop the numerical method for the research on the MPS with the PMF. 
They make clear that these PMFs raise the MPS on the smaller scales.

Furthermore, nonlinear effects of a PMF have been studied. References \cite{Jedamzik:1996wp,Subramanian:1997gi,Mack:2001gc} establish a damping scale of a PMF, and Ref. \cite{2012PhRvD..86d3510S} makes clear that the nonlinear effect from Alfven wave drops the MPS on the wave number $k~>~1.0~\mathrm{h/Mpc}$.


An average of a PMF strength as the background in the Universe can be assumed to be zero, while an ensemble average of a PMF energy density $\rho_\mathrm{MF}$, which is a zero order source from the PMF in the linear perturbation theory, in the Universe is not zero.
Recently, Refs. \cite{2012PhRvD..86l3006Y,2014PhRvD..90j3001Y} constrain $\rho_\mathrm{MF}$ by the light element abundances up to Li produced in the big bang nucleosynthesis, and they report the ratio of the ensemble average of the PMF energy density to the total photon energy density is less than 0.2. In the precision cosmology with the recent CMB observations, this value is too large to ignore. Therefore, to accurately study the physical processes in the precision cosmology, several authors study the effects of zero order sources from a PMF on the cosmology in the linear perturbation theory.

Ref. \cite{Adams:1996cq} assumes that a PMF is homogenous and the direction of the PMF is fixed, indicating that the magnetosonic wave from a PMF distorts the amplitude of the temperature fluctuations of the CMB.
Reference \cite{Kahniashvili:2006hy} researches the effects of a magnetic sonic wave for a stochastic magnetic field on the CMB analytically. Reference \cite{Koh:2000qw} studies the effects of an energy density of a PMF on the CMB. 
Also, Ref. \cite{2014PhRvD..89j3528Y} numerically analyzes teh effects of zero order sources from a stochastic isotropic and homogenous PMF on the CMB in the linear perturbation theory for the first time. This study indicates that the overall effects of an ensemble average energy density and a magnetosonic wave of from a PMF as the zero order sources change the peak positions of the temperature fluctuations of the CMB and decrease the amplitude of the CMB around the first peak.

The peak position of the MPS is determined by the horizon scale at the matter-radiation equality time $\tau_\mathrm{eq}$, and $\tau_\mathrm{eq}$ is determined by the ratio of the total matter density ($\rho_\mathrm{M}$) to the total radiation energy density ($\rho_\mathrm{R}$).
$\rho_\mathrm{R}$ is proportional to $a^{-4}$, where $a$ is the scale factor, and a PMF energy density $\rho_\mathrm{MF}$ is also proportional to $a^{-4}$. Therefore, if there is a sufficiently large $\rho_\mathrm{MF}$ in the radiation-dominated era, we should include $\rho_\mathrm{MF}$ in $\rho_\mathrm{R}$.
Thus, ones expect that the presence of the PMF increases the total radiation (like) energy density and delays the matter-radiation equality time, and, finally, the PMF changes the peak position of the MPS.

In Ref. \cite{2014PhRvD..89j3528Y}, also, the sound speed of the photon-baryon fluid is raised by a magnetosonic wave from a PMF as the zero order sources in the linear perturbation theory. A large sound speed causes strong pressure and suppresses the time evolution of the matter density field.
Therefore, if the magnetosonic wave from the PMF is large enough, ones expect that the amplitude of the MPS decreases.

In this paper, we consider the ensemble average of the PMF energy density and the magnetosonic wave from the PMF, which are the zero order sources of the linear perturbation theory, to theoretical computations of the MPS for the first time. 
We also analyze the effects of both of the zero and first order sources from the PMF on the MPS.

In this paper, we use the best-fit cosmological parameters determined by the observation data sets of the Planck and WMAP 9yr Collaborations \cite{Planck_I_Arxiv,WMAP_9yr_Arxiv}.
We modify the CAMB code \cite{camb} for computing the MPS with the power law (PL) PMF. The reduced Planck constant ($\hslash$) and the speed of light (c) are a unit.
\section{\label{sec:models}Models}
In this section, we will mention the models of the PMF in this paper.

A minimum scale of a PMF at last scattering is derived by
\begin{eqnarray}
L_* \sim 3.68 \times 10^{-15}
\left(\frac{\eta_0}{\eta}\right)^{\frac{1}{2}}
\left(\frac{t}{t_*}\right)^{\frac{1}{2}}
\mathrm{Mpc} 
\end{eqnarray}
\cite{Grasso:2000wj,Dendy:1990booka},
where $\eta$ is the baryon to photon ratio ($\eta_0 = 6.19\times 10^{-10}$\cite{2013ApJS..208...19H}) and 
$t$ is the age of the Universe ($t_* =3.76 \times 10^{5}$ year \cite{2013ApJS..208...19H}).
A PMF on the scale much bigger than $L_*$ is hard to dissipate by $t_*$, and such a PMF is "frozen in" the dominant fluids \cite{Dendy:1990booka}.
This scale is much smaller than the ranges for the linear perturbation theory. 
Therefore, we assume that a PMF for the linear perturbation theory is frozen in.
We also assume that a PMF is generated before the matter-radiation equality. 
We assume that the PMF is statically homogeneous, isotropic, and random. In this case, the ensemble average of the PMF strength is zero. On the other hand, the ensemble average of the PMF energy density is finite.
We consider the effects of the ensemble average energy density and the magnetosonic wave of the PMF on the MPS, numerically. 
\subsection{\label{sec:models}The energy density and the sound speed with the PL-PMF model }
We use the PL-PMF model \cite{Mack:2001gc}. 
This model assumes that a PMF is generated in the inflation era.
A spectrum and a two-point correlation function of the PL-PMF model are given by \cite{Mack:2001gc}
\begin{eqnarray}
\langle B(k)B^\ast(k)\rangle  \propto k^{n_\mathrm{B}}, \\
\left\langle B^{i}(\mbi{k}) {B^{j}}^*(\mbi{k}')\right\rangle 
	=	({(2\pi)^{n_\mathrm{B}+8}}/{2k_\lambda^{n_\mathrm{B}+3}})
		\left[
{B^2_{\lambda}}/{\Gamma\left(\frac{n_\mathrm{B}+3}{2}\right)}
\right]
		k^{n_\mathrm{B}}P^{ij}(k)\delta(\mbi{k}-\mbi{k}'),~
                k < k_\mathrm{max},
		\label{two_point1} 
\end{eqnarray}
where 
$k$ is the wave number, 
$n_\mathrm{B}$ is the spectrum law of the PMF, 
$B_\lambda=|\mathbf{B}_\lambda|$ is the comoving field strength by smoothing over a Gaussian sphere of radius $\lambda=1$ Mpc ($k_\lambda = 2\pi/\lambda$),
$\Gamma(x)$ is the gamma function, 
$i$ and $j$ are the spatial indices and the integer numbers [$\in$ (1, 2, 3)],
$P^{ij}(k)=\delta^{ij}-\frac{k{}^{i}k{}^{j}}{k{}^2}$,
$k_\mathrm{max}$ is the cutoff scale and is defined by the PMF damping \cite{Jedamzik:1996wp, Subramanian:1997gi,Mack:2001gc}, 
and 
$B^{i}(\mbi{k})$ is from the Fourier transform convention:
$B^{i}(\mbi{k}) = \int d^3 x e^{i\mbi{k}\cdot\mbi{x}} B^{i}(\mbi{x})$.
Here $B(a,\mbi{x}) = B(\mbi{x})/a^2$, where $a$ is the scale factor.

In this case, the ensemble average of the PMF energy density $\rho_\mathrm{MF}$
is finite and this value on the physical field is given by \cite{Subramanian:1997gi,Subramanian:1998fn}
\begin{eqnarray}
\rho_\mathrm{MF}
= 
\frac{1}{8\pi a^4}
\frac{
B^2_\lambda
}
{
   \Gamma
   \left(
      \frac{n_\mathrm{B}+5}{2}
   \right)
}
\left[
(\lambda k_\mathrm{max})^{n_\mathrm{B}+3}
-
(\lambda k_\mathrm{min})^{n_\mathrm{B}+3}
\right].
\label{eq:BG_PL_PMF_energy_density}
\end{eqnarray}
where 
$k_\mathrm{min}$ is the minimum wave numbers
\footnote{The minimum wave number of the PMF damping in this paper (at $B_\lambda$ = 10 nG and $n_\mathrm{B}$ = 0.0) is of the order of 1 Mpc$^{-1}$. Therefore it is sufficiently larger than $k_\mathrm{eq}$ which is of the order of 0.01 Mpc$^{-1}$. Here $k_\mathrm{eq}$ is determined by the horizon scale at the equality time.},
Since the strength of the frozen-in PMF on the physical field is proportional to $a^{-2}$, $\rho_\mathrm{MF}$ is proportional to $a^{-4}$.
If some vorticity anisotropies of an inflationary origin generate a PMF, $k_\mathrm{[min]}/k_\mathrm{max}$ is assumed to be very small.
In this case, the last term in Eq. (\ref{eq:BG_PL_PMF_energy_density}) is negligible, and we obtain:
\begin{eqnarray}
\rho_\mathrm{MF}
\sim
\frac{1}{8\pi a^4}
\frac{
B^2_\lambda
}
{
   \Gamma
   \left(
      \frac{n_\mathrm{B}+5}{2}
   \right)
}
(\lambda k_\mathrm{max})^{n_\mathrm{B}+3}.
\label{eq:BG_PL_PMF_energy_densityII}
\end{eqnarray}

The scale factor at the equality time is determined by the matter-to-radiation ratio as follows:
\begin{eqnarray}
a_\mathrm{eq} = \frac{\rho_\mathrm{R}}{\rho_\mathrm {M}},\label{eq:eq}
\end{eqnarray} 
where $\rho_\mathrm{R}$ and $\rho_\mathrm{M}$ are the total radiation(like) energy density and the total matter density today $ (a=1)$, respectively.
If $\rho_\mathrm{M}$ is fixed, this equation shows that it is important for the total radiation(like) energy density to determine the equality time. Thus, the peak position of the MPS is dominated by $\rho_\mathrm{R}$.
Since the time evolution of the ensemble average of the PMF energy density $\rho_\mathrm{MF}$ is proportional to $a^{-4}$ in the same way as the radiation energy densities, the presence of the PMF in the radiation-dominated epoch delays the matter-radiation equality time $a_\mathrm{eq}$. 
The peak of the MPS is determined by the horizon scale at $a_\mathrm{eq}$. 
The ensemble average energy density of the PMF is not a first order perturbation but a zero order factor (non-perturbed factor) for the linear perturbation equations.
Therefore, $\rho_\mathrm{MF}$ as a zero order factor for the linear perturbation theory gives an important effect on the MPS.

The effective sound speed taking into account the PMF, $c_\mathrm{sA}$, is introduced by \cite{Adams:1996cq,Kahniashvili:2006hy}. 
We use the expression of $c_\mathrm{sA}$ as follows \cite{2014PhRvD..89j3528Y}\footnote{
Reference \cite{Adams:1996cq} assumes that the direction of the magnetic field is fixed; in other words, this field is anisotropic and introduces the sound velocity with the Alfven velocity by $c^2_\mathrm{sA} = c^2_s + c^2_\mathrm{A}\sin^2\theta $. 
On the other hand, we assume the PMF is stochastic isotropic. In a statistical cosmological study, we are interested not in local magnetic effects, but in global ones, which are average universewide. 
Therefore, we need the average of the second term in $c^2_\mathrm{sA}$ per $\theta$, and it becomes $\frac{1}{2} c^2_\mathrm{A}$.}
\begin{eqnarray}
c^2_\mathrm{sA}
=
c^2_\mathrm{s}
+
\frac{1}{2}
c^2_\mathrm{A}.
\label{result_c_sA}
\end{eqnarray}
Here,
$c_\mathrm{s}$ is the sound velocity of a fluid without a PMF and given by $c^2_\mathrm{s} = c^2_\mathrm{b\gamma} = 1/\{ 3(1+\frac{3\rho_\mathrm{b}}{4\rho_\gamma}) \} = 1/\{ 3\left(1+R\right) \} $, 
where $\rho_\mathrm{b}$ and $\rho_\gamma$ are the baryon density and the photon energy density, and $R$ is $\frac{3}{4}\frac{\rho_\mathrm{b}}{\rho_\gamma}$. 
$c_\mathrm{A}$ is the Alfven velocity in the baryon-photon fluid with a PMF and it is  proportional to $\rho_\mathrm{MF}$, which is a zero order factor for the linear perturbation theory, as follows:
$c^2_\mathrm{A} = \frac{2\rho_\mathrm{MF} }{\rho_\gamma + \rho_\mathrm{b}}$ \cite{2014PhRvD..89j3528Y}. 

$ c^2_\mathrm{sA} $, which is the sound velocity with the PMF is not an effective factor for the linear perturbation equations at the subhorizon and superhorizon. Therefore, to estimate the initial conditions with the PMF, we do not have to consider $c^2_\mathrm{sA}$, and it is only necessary to consider $\rho_\mathrm{MF}$ for the Friedmann equation and the conservation of energy momentum tensor.
Therefore, we just have to change the total energy density, $\rho$ without $\rho_\mathrm{MF}$ to this value with $\rho_\mathrm{MF}$, e.g., $\rho = \rho_\mathrm{R} + \rho_\mathrm{M}$ to $\rho = \rho_\mathrm{R} + \rho_\mathrm{M} + \rho_\mathrm{MF}$. 
Finally, the expressions of the initial conditions with the PMF are  not different from ones in Refs. \cite{2008PhRvD..77d3005Y,Shaw:2009nf}.
 
In this study, we assume that the PMF does not correlate with the primary perturbation.
In this case, we do not need to take into account the correlation term between the PMF and the primary in Eq. (22) of Ref. \cite{2014PhRvD..89j3528Y}. 
\section{\label{sec:Rs}Results and Discussions}
The ensemble average energy of the PMF $\rho_\mathrm{MF}$ and the effective sound speed with the PMF $c_\mathrm{sA}$ are not the first order perturbed factors but the nonperturbed sources (the zero order sources) in the linear perturbation equations for the cosmology. 
To avoid redundant representations, we will call these sources "the zero order sources of the PMF" in this section.

On the other hand, the PMF source whose order is equivalent to the first order perturbed factor in the linear perturbation equations for the cosmology [e.g., the last term of Eq.(32) in Ref. \cite{2014PhRvD..89j3528Y}]  has been mainly analyzed in previous studies. 
For the avoidance of confusion between this source and "the zero order sources of the PMF", we will call it "the first order sources of the PMF" in this section.

The evolution rate of the perturbation of the matter density fields is proportional to $\sqrt{G\rho_m}$, where $G$ is Newton's constant. The expansion rate of the Universe on the radiation-dominated era is proportional to $\sqrt{G\rho_r}$. 
On the radiation-dominated era, the total energy density of the radiations is larger than the total matter density. Therefore, $\sqrt{G\rho_m}$ on the radiation-dominated era is less than $\sqrt{G\rho_r}$, and the perturbations of the matter density fields cannot be evolved in the radiation-dominated era. 
This is called the {\it Meszaros effect} \cite{1974A&A....37..225M}.
After the equality time, the matter density fields can be affected by the potential and start to evolve in the horizon. Also, on the superhorizon, the matter perturbations can evolve without the Meszaros effect.

The evolution of the matter perturbations with the Meszaros effect is summarized as follows: In the radiation-dominated era (1) the matter perturbations in the superhorizon can evolve and (2) once the matter perturbations enter the horizon, they do not grow. 

Small scales enter the horizon scale earlier, and the matter perturbations on the longer wave numbers are influenced by the Meszaros effect for a longer time. Therefore, the amplitudes of the MPS on the longer wave numbers are more strongly damped by the Meszaros effect.

From Ref, \cite{Ma:1995ey}, the potentials decay on the horizon in the radiation-dominated era is given by
\begin{eqnarray}
\phi \propto
\sqrt{\frac{1}{k^3\tau^3 c^3_\mathrm{S}}}
J_{3/2}\left(k\tau c_\mathrm{S}\right),
\label{eq_decay_potential}
\end{eqnarray}
where $J_n$ is Bessel functions of the first kind, 
$\tau$ is the conformal time and defined by $\tau = \int^t_0 \frac{1}{a(t')}dt'$,
and $c_\mathrm{S}$ is the sound speed and corresponds to $\sqrt{1/3}$ without a PMF.
This equation shows that the potentials which enter the horizon on the radiation-dominated era are affected by the suppression of the sound speed. Since the matter density perturbations are influenced by the potentials, the amplitudes of the matter density perturbations are also suppressed.

As a result, both of the above-mentioned effects damp the amplitudes of the MPS on $k > k_\mathrm{eq}$, where $k_\mathrm{eq}$ is determined by the horizon scale at the equality time $\tau_\mathrm{eq}$. On the other hand, the wave number modes that enter the horizon scales after the equality time are not dominated by these effects, and the MPS amplitudes on $k < k_\mathrm{eq}$ are not damped.
Thus, one can understand that the MPS with the Meszaros effect and the potential damping from the sound speed has the peak at $k_\mathrm{eq}$. 

As mentioned Sec. II, the equality time with $\rho_\mathrm{MF}$ is later than without $\rho_\mathrm{MF}$. 
The values of these damping are dependent on how long the correspondent wave number modes are on the horizon for the radiation-dominated era. From Eq. (\ref{eq_decay_potential}), the larger sound speed also suppresses the amplitude of the MPS more strongly.
Furthermore, the horizon scale at the equality time with $\rho_\mathrm{MF}$ becomes larger than without $\rho_\mathrm{MF}$. Finally, the peak position of the MPS with $\rho_\mathrm{MF}$ shifts to smaller wave numbers.

In case of the later equality time, since the effective term of the Meszaros and the potential damping effects is longer, the amplitudes of the MPS on wave numbers larger than the peak position become smaller. The sound speed with the PMF becomes larger than without the PMF and suppresses the MPS amplitudes on wave numbers larger than the peak positions.

\begin{figure}
\includegraphics[width=1.0\textwidth]{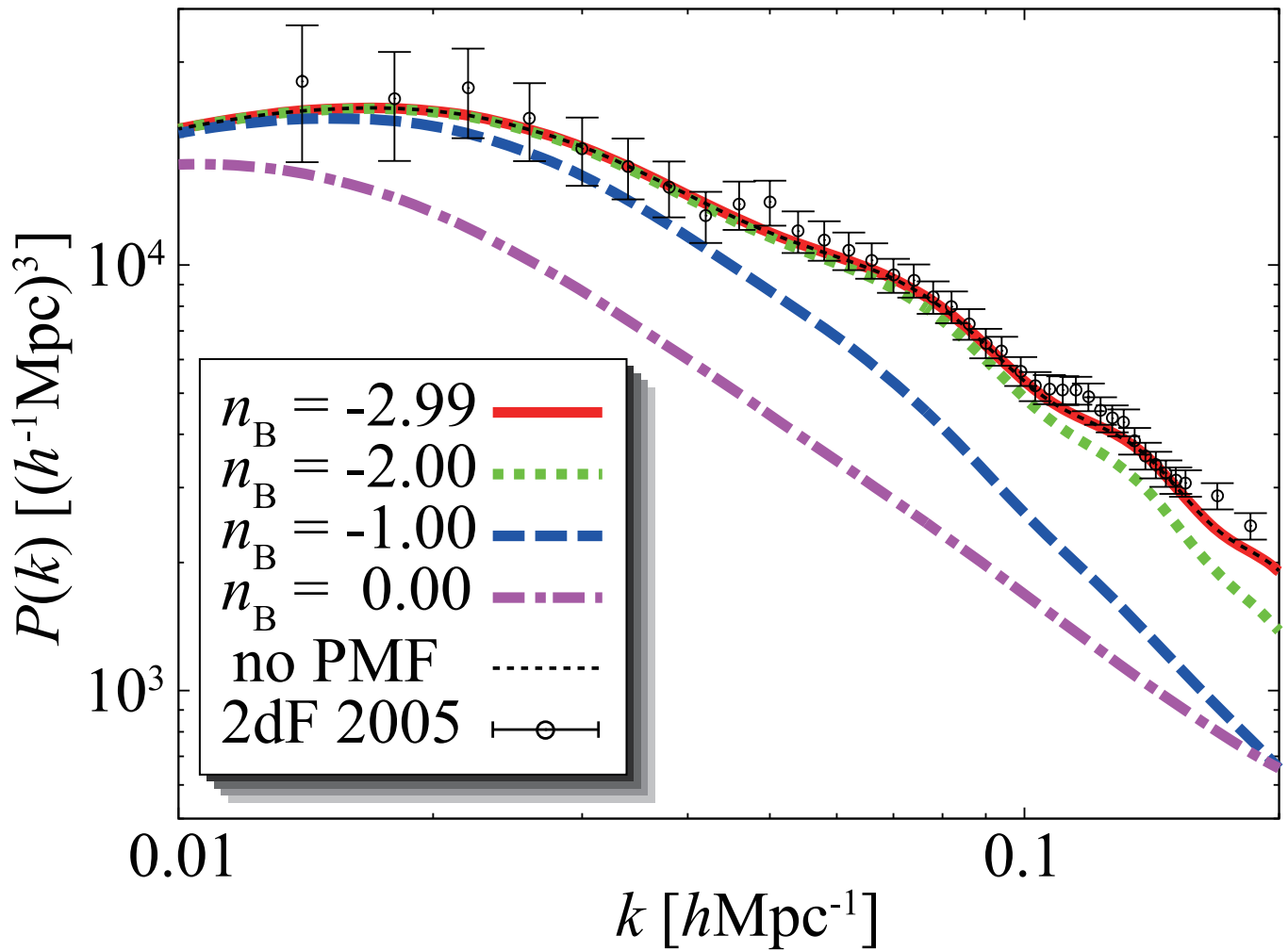}
\caption{\label{fig1}
The effects of the PMF on the MPS in the case of the fixed PMF strength as $B_\lambda = 10$ nG.
The thin dotted curve in this figure is the theoretical result from the Planck and WMAP 9 yr best-fit parameter in the $\Lambda$CDM model \cite{Planck_I_Arxiv,WMAP_9yr_Arxiv}.
The dots with error bars show 2dF \cite{Cole:2005sx}.
The solid, dotted, dashed, and dash-dotted curves are the theoretical result with the PMF effects of $(n_\mathrm{B},~\rho_\mathrm{MF}/\rho_\gamma) = $ $(-2.99,~1.00\times 10^{-5})$, $(-2.00,~5.84 \times 10^{-4})$, $(-1.00,~9.65 \times 10^{-3})$, and $(0.00,~8.63 \times 10^{-2})$, respectively.
} 
\end{figure}

Figure \ref{fig1} shows the MPS with the PMF. We can see that the peak positions of the MPS with the PMF shift to the lower wave numbers, and the amplitudes of the MPS with the PMF are also lower than ones without the PMF (no PMF) on wave numbers larger than the peak position.
The PMF damps the MPS on the wave number range of $0.01 h\mathrm{Mpc}^{-1}< k <~0.2~h\mathrm{Mpc}^{-1}$, where the observational data points in Fig. \ref{fig1} are.
If a PMF strength with higher $n_\mathrm{B}$ is of the order of 1 nG, which is expected by the observations of the magnetic field in the cluster of galaxies \cite{2012ApJ...747L..14V,2012A&ARv..20...54F}, such a PMF critically affects the MPS as shown by the dotted ($n_\mathrm{B}$ = -2.0), dashed ($n_\mathrm{B}$ = -1.0), and dash-dotted  ($n_\mathrm{B}$ = 0.0) curves in Fig. \ref{fig1}. Therefore, we expect the PMF with the higher $n_\mathrm{B}$ can be strongly constrained by the observational MPS data sets.

\begin{figure}
\includegraphics[width=1.0\textwidth]{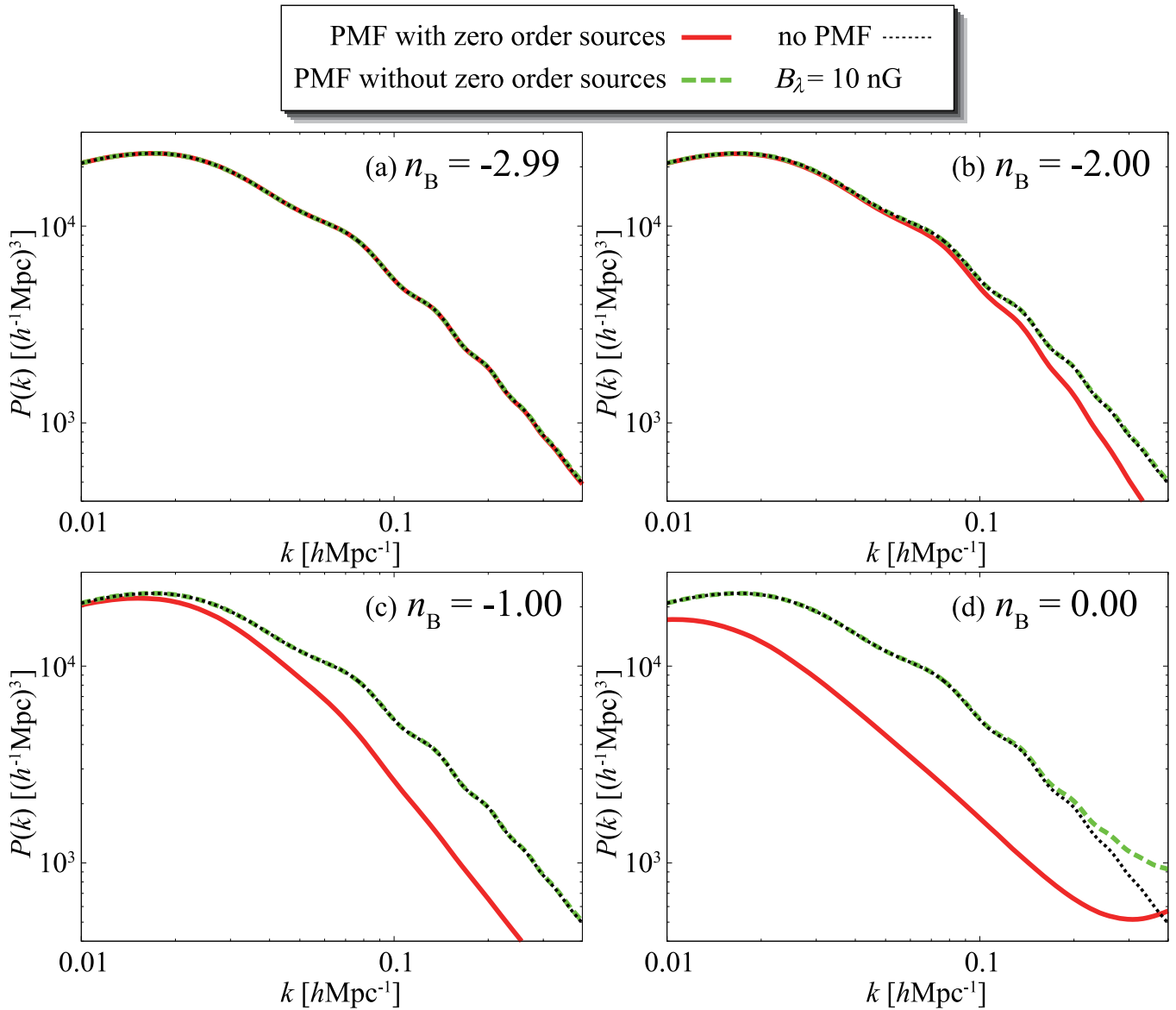}
\caption{\label{fig2}
The effects of the zero and first order sources of the PMF on the MPS in the case of the fixed PMF strength as $B_\lambda = 10$ nG.
The thin dotted curves in this figure are the theoretical result from the Planck and WMAP 9 yr best-fit parameter in the $\Lambda$CDM model \cite{Planck_I_Arxiv,WMAP_9yr_Arxiv}.
The dashed curves show the MPS with the first order sources and without the zero order sources from PMF. The solid curves show the MPS with the first and the zero order sources from PMF.
The PMF parameters sets in the panels (a), (b), (c), and (d) are $(n_\mathrm{B},~\rho_\mathrm{MF}/\rho_\gamma) = $ $(-2.99,~1.00\times 10^{-5})$, $(-2.00,~5.84 \times 10^{-4})$, $(-1.00,~9.65 \times 10^{-3})$, and $(0.00,~8.63 \times 10^{-2})$, respectively.
} 
\end{figure}

\begin{figure}
\includegraphics[width=1.0\textwidth]{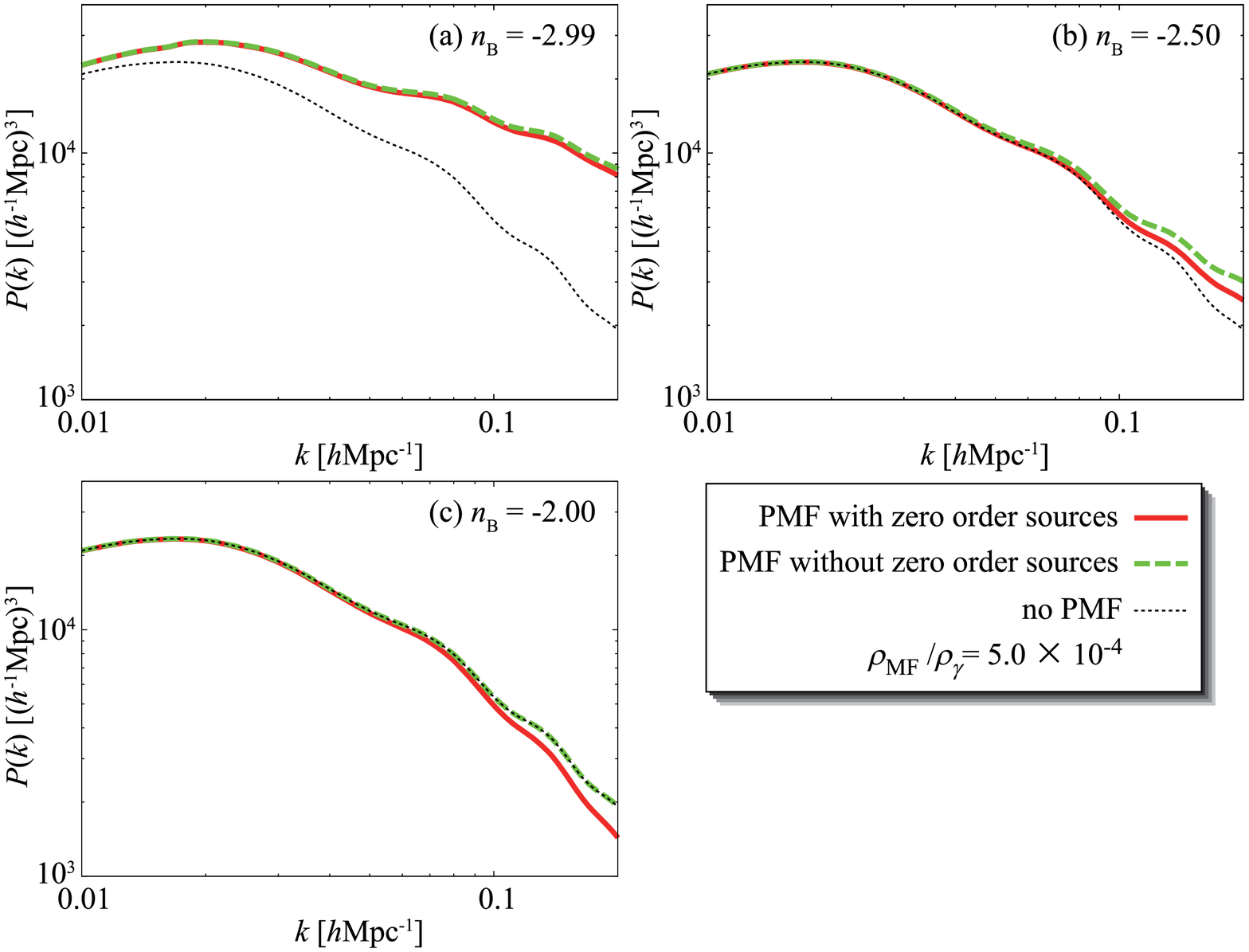}
\caption{\label{fig3}
The effects of the PMF on the MPS in the case of $\rho_\mathrm{MF}/\rho_\gamma = 5.0 \times 10^{-4}$.
The thin dotted curves in this figure are the theoretical result from the Planck and WMAP 9 yr best-fit parameter in the $\Lambda$CDM model \cite{Planck_I_Arxiv,WMAP_9yr_Arxiv}.
The solid curves in panels (a), (b), and (c) show the MPS with the zero and first order sources from the PMF. The dash-dotted curves in panels (a), (b), and (c) show the MPS with first order sources from the PMF and without zero order sources from the PMF.
The PMF parameters in panels (a), (b), and (c) are $(n_\mathrm{B},~B_\gamma) = $ $(-2.99,~ 70.9 \mathrm{nG})$, $(-2.50, ~ 25.0 \mathrm{nG})$, and $(-2.00, ~ 9.11 \mathrm{nG})$, respectively.
} 
\end{figure}

To investigate features of the zero order and the first order sources from the PMF effects on the MPS on each PMF parameter, we will analyze two cases: the first case is fixing a PMF strength of 10 nG (Fig. \ref{fig2}), and the second is fixing a ratio of PMF-to-photon energy densities of 5.0 $\times 10^{-3}$ (Fig. \ref{fig3}).

In case of the fixed PMF strength, the amplitudes of the MPS with the first order sources from the PMF on larger wave numbers positively correlate with the power law indexes $n_\mathrm{B}$ \cite{2006PhRvD..74l3518Y,2012PhR...517..141Y}.
Therefore, considering the effects of the first order sources from the PMF only, these effects of relatively small $n_\mathrm{B}$ on the MPS are very small as the panels (a) - (c) of Fig. \ref{fig2} indicates. However, in fact, the MPS is sufficiently suppressed by the zero order sources from the PMF except $n_\mathrm{B} = -2.99$ as the panels (b) - (d) of Fig. \ref{fig2} indicates.

Panel (a) of Fig. \ref{fig3} shows the MPS with the zero and first order sources from the PMF of $\rho_\mathrm{MF}/\rho_\gamma = 5 \times 10^{-4}$.
On sufficiently small $n_\mathrm{B}$, the first order sources from the PMF dominate the overall PMF effects on the MPS [panel (a) of Fig. \ref{fig3}].
In case of fixing $\rho_\mathrm{MF}$, the amplitude of the spectrum from the first order sources from the PMF has a negative-correlation with the PMF power law index $(n_\mathrm{B})$ as Eq. \ref{eq:BG_PL_PMF_energy_densityII}, and we find that the MPS amplitudes of sufficiently small $n_\mathrm{B}$ are extremely raised by the first order sources from the PMF. 
In this case, from panel (a) of Fig. \ref{fig3}, the effects of the zero order sources from the PMF on the MPS are much smaller than the effects of the first order sources from the PMF. 
Thus, the overall effects of the PMF on the MPS become slightly weaker than in previous studies without the zero order sources from the PMF. 

On relatively large $n_\mathrm{B}$ as panel (b) of Fig. \ref{fig3} indicates, we find that the first order sources from the PMF raise the MPS on the larger wave numbers slightly. 
On the other hand, the zero order sources from the PMF reduce the addition of the MPS from the first order sources from the PMF.
In this case, finally, 
the overall effects of the PMF become weaker than in previous studies without the zero order sources from the PMF.

On sufficiently large $n_\mathrm{B}$, the effects of the first order sources from the PMF are very small and the MPS with the PMF is only suppressed by the effects of the zero order sources from the PMF as the panel (c) of Fig. \ref{fig3} indicates.
The curve of the MPS with the zero order sources from the PMF in the panel (c) of Fig. \ref{fig3} corresponds approximately to the MPS with the pure zero order sources from the PMF and without the first order sources from the PMF. 

As a conclusion of these analyses, to research the PMF effects on the MPS on the wider wave numbers range correctly, we need to consider both effects of the zero and first order sources from PMF simultaneously.

In case of considering the overall effects of the PMF on the MPS correctly, we discuss how we constrain the PMF parameters $(B_\lambda, n_\mathrm{B})$ from the MPS observations.
If $n_\mathrm{B}$ is nearby -3.0, as mentioned above and shown in Figs \ref{fig2} and \ref{fig3}, the first order sources from the PMF dominate the overall effects of the PMF on the MPS. Therefore, we expect that constraints on the PMF parameters with effects of the zero order sources from the PMF on the MPS are not much different from the previous studies.

When $n_\mathrm{B}$ is relatively large, the zero order sources from the PMF suppress additions of the MPS amplitudes from the first order PMF sources as shown in panel (b) of Fig. \ref{fig3}, and the overall PMF effects are weaker than the effects of the pure first order sources from the PMF as previous studies show. Thus, we expect that the constrained strengths of the PMF on ranges of relatively large $n_\mathrm{B}$ become larger than in previous studies.

In case of sufficiently large $n_\mathrm{B}$, the zero order sources from the PMF dominate the overall effects of the PMF on the MPS. From Figs. \ref{fig1}-\ref{fig3}, the MPS with the effects of the zero order sources from the PMF of $B_\lambda = 10$ nG, which corresponds to the upper limits of the expected PMF strength by the observations in the clusters of the galaxies, is strongly suppressed in the linear-perturbation theory region. In this case, differences in MPS with and without the overall PMF are much larger than differences in MPS with and without the first order sources from the PMF. Thus, if we consider the PMF effects correctly, we expect that the constrained PMF strengths on sufficiently large $n_\mathrm{B}$ become much smaller than the results without the zero order sources from the PMF.

In summary, we investigate the overall effects of the zero and first order sources from the PMF on the MPS.
Except on $n_\mathrm{B}$ around -3.0, we find that the suppression effects of the zero order sources from the PMF are not neglected for studying the MPS with the PMF. 
We also mention that the constrained PMF strengths with the correct effects of the overall PMF are different from the results without the zero order sources from the PMF. 

When we estimate the weak lensing effects of the CMB, the MPS is very important.
The CMB polarization gives us useful information to research the early Universe, e.g., the inflation theory and the background gravitational wave, and is affected by the weak lensing effect too. Recently, many authors discuss the observations of the CMB polarization in the precision cosmology. When we research and correctly constrain the cosmological parameters with the PMF from the CMB including the polarization, we should consider the effects of the zero order sources from the PMF as our work.

\begin{acknowledgments}
This work has been supported in part by Grants-in-Aid for Scientific Research 
(Grant No. 25871055) of the Ministry of Education, Culture, Sports,
Science and Technology of Japan.
We are grateful to Yolande McLean for improving the English in this paper.
\end{acknowledgments}
\bibliographystyle{apsrev}

\end{document}